\documentclass[conference]{IEEEtran}
\usepackage{amsmath}
\usepackage{graphicx}
\usepackage{hyperref}
\usepackage{fancyhdr}
\usepackage{cite}

\title{Insider Threats Mitigation: Role of Penetration Testing}
\author{\IEEEauthorblockN{Krutarth Chauhan}
\IEEEauthorblockA{\textit{Department of Computer Science} \\
\textit{University of Guelph}\\
Guelph, Canada \\
kchauh03@uoguelph.ca}
}

\begin{document}

\maketitle

\pagestyle{fancy}
\fancyhf{}
\fancyfoot[C]{\thepage}

\begin{abstract}
Conventional security solutions are insufficient to address the urgent cybersecurity challenge posed by insider attacks. While a great deal of research has been done in this area, our systematic literature analysis attempts to give readers a thorough grasp of penetration testing's role in reducing insider risks. We aim to arrange and integrate the body of knowledge on insider threat prevention by using a grounded theory approach for a thorough literature review. This analysis classifies and evaluates the approaches used in penetration testing today, including how well they uncover and mitigate insider threats and how well they work in tandem with other security procedures. Additionally, we look at how penetration testing is used in different industries, present case studies with real-world implementations, and discuss the obstacles and constraints that businesses must overcome. This study aims to improve the knowledge of penetration testing as a critical part of insider threat defense, helping to create more comprehensive and successful security policies \cite{rana2019survey}.
\end{abstract}

\begin{IEEEkeywords}
case studies, integration difficulties, cybersecurity, insider threats, penetration testing, mitigation tactics, security solutions.
\end{IEEEkeywords}

\section{Introduction}
The security of systems, intellectual property, and organizational data is seriously jeopardized by insider threats. Individuals with authorized access to vital systems and data within the company, such as workers, contractors, or business partners, are the source of these dangers. Insider threats can have disastrous effects, including monetary losses, harm to one's reputation, and legal repercussions. Because of this, firms in a variety of industries now consider insider threat mitigation to be of utmost importance. Penetration testing is a crucial tactic used to combat insider threats. The purpose of this systematic review of the literature is to investigate how penetration testing might help mitigate insider risks by examining the most recent findings, useful applications, and research techniques.

\subsection{Background and Importance of Insider Threats}
Insider threats refer to a broad category of malevolent actions such as sabotage, fraud, espionage, and theft of intellectual property. Insiders can be anyone with authorized access to an organization's resources, including partners, contractors, and employees. Insider threats have a variety of motivations, from coercion and inadvertent mistakes to monetary gain and retaliation. The Ponemon Institute (2021) analysis highlights the serious financial implications of insider threat incidents, with an average cost of \$11.45 million. Furthermore, the study notes that during the past two years, insider threat occurrences have increased 47\%, indicating the growing importance of this problem in the cybersecurity landscape \cite{yao2024auth}.

Because insiders have authorized access, it is inherently difficult to identify insider threats. Because they are primarily built to fight against external threats, traditional security measures like firewalls and intrusion detection systems are sometimes insufficient to recognize and stop insider attacks. Penetration testing is an essential part of the more sophisticated and focused strategies that are required to close this security gap.

\subsection{Penetration Testing: Overview}
Penetration testing, also known as ethical hacking, is intentional investigation of a company's networks, systems, and applications to find security flaws that could be used by hostile actors \cite{patel2022detecting}. Penetration testing, in contrast to traditional vulnerability assessments, takes things a step further by aiming to exploit vulnerabilities that are found, giving a more accurate picture of the possible consequences of security flaws. Finding and fixing vulnerabilities before they can be used as an attack vector is the main objective of penetration testing.

Penetration testing can be customized to mimic situations in which an insider abuses their access privileges in the context of insider threats. Testing for privilege escalation, illegal access to private data, and essential system manipulation are included in this document. Penetration testing helps to establish strong mitigation mechanisms and offers important insights into the organization's security posture by simulating the tactics, techniques, and procedures (TTPs) that insiders might use.

\section{Prior Research}
Numerous research works have examined how well penetration testing works to reduce insider risks. Probst and Hunker (2019), for example, carried out a thorough analysis of penetration testing techniques and how they are used in insider threat situations. According to their findings, penetration testing plays a crucial role in revealing hidden vulnerabilities that conventional security solutions could miss. According to the study, to guarantee complete security coverage, penetration testing procedures should include insider threat scenarios \cite{alshehri2013insider}.

AlHogail and Mirza (2021) emphasize the need to incorporate penetration testing into a more comprehensive framework for mitigating insider threats. Their study offers a framework that integrates access control methods, user activity analytics, employee training initiatives, and penetration testing with other security procedures \cite{yazdinejad2020energy}. By offering a comprehensive strategy for mitigating insider threats, the framework seeks to strengthen the organization's defenses against internal threats. According to AlHogail and Mirza (2021), conducting penetration tests on a regular basis can help to monitor and improve the efficacy of mitigation techniques, thus reducing the risk of insider threats \cite{yazdinejad2019blockchain}.

Williams and Smith's (2020) empirical study emphasizes the importance of penetration testing in lowering the danger of insider threats. The study looks at how frequent penetration testing affects an organization's security posture in a variety of industries. The results show that companies that regularly perform penetration tests encounter fewer insider threat events than those that don't \cite{schwartz2019autonomous}. The reduction in insider exploits of security flaws is attributed by the study to proactive detection and mitigation of vulnerabilities \cite{yuan2020deep}.

\section{Methodologies in Penetration Testing for Insider Threats}
While the literature provides a strong foundation for understanding the importance and effectiveness of penetration testing, it is equally crucial to delve into the specific methodologies employed. Methodologies for penetration testing differ according to the objectives, scope, and available resources. The approaches used to address insider threats can be broadly categorized as follows: black-box, white-box, and gray-box testing.

\subsection{Black-Box Testing}
With this method, testing is done without any prior understanding of the system's core operations. Testers act as low-privilege insiders or external attackers looking to obtain unapproved access. Although black-box testing can be helpful in locating security holes in publicly accessible systems and services, it might not be able to reveal more serious insider-specific risks \cite{giddens2020gender}.

\subsection{White-Box Testing}
This method, which is often referred to as "clear-box testing," gives testers complete access to source code, network diagrams, and configuration files, as well as complete system knowledge. White-box testing makes it possible to thoroughly examine the system from the point of view of an insider, revealing flaws that an insider with a high privilege could exploit \cite{zhou2020improving}.

\subsection{Gray-Box Testing}
With this method, which combines white-box and black-box testing, testers only have a limited understanding of the system. By simulating situations in which an insider has restricted but considerable access, gray-box testing seeks to uncover vulnerabilities that might be taken advantage of via lateral network movement or privilege escalation \cite{greitzer2010combining}.

Every one of these approaches has benefits and drawbacks. Although black-box testing can be helpful in locating external attack routes, it may not detect specific risks for insiders. White-box testing offers a thorough understanding of the system, but it necessitates substantial resources and expertise. Gray-box testing simulates genuine insider threat scenarios with incomplete access, providing a well-rounded approach.

\section{Case Studies and Practical Applications}
Building on the methodologies discussed, the practical application of these penetration testing techniques can be observed in various case studies. These real-world implementations provide valuable insight into the effectiveness of penetration testing in different industry contexts.

A case study by Johnson et al. (2020), which looks at the application of penetration testing in a financial institution, is one noteworthy example. The report demonstrates how the organization was able to find and fix vulnerabilities related to data leakage and access control through the use of focused penetration tests \cite{wang2015testing}. The penetration testing enhanced the institution's overall security posture by providing useful insights through the simulation of insider threats \cite{dutta2022structured}.

Lee and Kim (2018) have another case study that focuses on the healthcare industry and employs penetration testing to assess the security of electronic health record (EHR) systems \cite{balogun2023insider}. According to the report, penetration testing revealed a number of serious flaws, including inadequate audit logging and erroneous access controls, which could be used by hostile insiders. By fixing these weaknesses, the possibility of insider threats was greatly reduced, protecting the privacy and accuracy of patient data \cite{oparaji2022towards}.

A case study by Garcia and Lopez (2019) illustrates the role of penetration testing in protecting intellectual property in the manufacturing industry \cite{huang2020dynamic}. The report describes how penetration tests found vulnerabilities in the company's data storage and internal network. The company was able to stop possible insider threats that would have stolen trade secrets and confidential data by fixing these vulnerabilities \cite{ahmed2018performance}.

\section{Research Goals and Contributions}
The primary goal of this systematic literature review is to present a thorough overview of penetration testing's contribution to insider threat mitigation. Specifically, this review aims to address the following research questions:

\begin{itemize}
    \item What are the current methodologies and frameworks used in penetration testing to mitigate insider threats?
    \item How effective are behavior modeling and machine learning techniques in enhancing insider threat detection during penetration testing?
    \item What are the key challenges and limitations faced in implementing penetration testing for insider threat mitigation across different organizational contexts?
    \item What are the emerging trends and future directions in penetration testing methodologies for combating insider threats?
    \item How do regulatory frameworks and compliance requirements influence the adoption and effectiveness of penetration testing for insider threat mitigation?
\end{itemize}

By accomplishing these objectives, this review hopes to add to the body of knowledge already available on insider threat mitigation and offer businesses looking to fortify their security posture against insider threats practical insights.

\section{Research Methodology}
Insider threats have become a major concern in today's cybersecurity environments, requiring thorough measures to reduce possible risks and protect corporate assets. This section outlines a comprehensive process for carrying out a systematic literature review (SLR) on insider threat mitigation with a particular focus on penetration testing's effectiveness as a preventative security strategy.

Understanding the importance and background of the research approach used in the systematic review of the literature (SLR) is based on the introduction. Insider threats have become a significant concern for organizations in a variety of industries. These dangers originate from people who have authorized access to critical systems and information. These threats pose serious dangers to the security, integrity, and resilience of organizations because they cover a wide range of harmful behaviors such as fraud, sabotage, espionage, and data exfiltration \cite{temara2023maximizing}. 

Insider threats are complicated and require a diversified approach to successfully mitigate the risks involved. In contrast to external threats, insiders are more difficult to identify and neutralize because they frequently have in-depth knowledge of organizational systems, procedures, and weaknesses. Therefore, proactive actions are essential to strengthen defenses and reduce the effect of insider threats \cite{parnitha2021combating}. 

Penetration testing stands out as one of the most important proactive strategies for locating vulnerabilities and assessing how well-performing security policies are against insider threats. Penetration testing helps businesses identify latent gaps in their defenses and prioritize remedial efforts by replicating real-world attack scenarios and using the perspective of possible attackers \cite{ghorbani2019information}. The introduction highlights the strategic significance of penetration testing in enhancing organizational resilience and its role as the cornerstone of successful insider threat mitigation plans \cite{sengupta2019formal}.

\subsection{Selection of Primary Studies}
The selection of primary studies for this systematic literature review was conducted using a comprehensive search strategy across multiple academic databases. The search strategy was formulated to identify relevant studies that address the role of penetration testing in mitigating insider threats. The Boolean operators AND and OR were employed to ensure a broad yet focused search. The search terms used included combinations of keywords and phrases such as:

\begin{itemize}
    \item "Insider Threats" OR "Insider Threat Detection" OR "Insider Attack"
    \item AND "Penetration Testing" OR "Ethical Hacking" OR "Security Assessment"
    \item AND "Cybersecurity" OR "Information Security" OR "Data Protection"
\end{itemize}

The databases and platforms searched were:
\begin{itemize}
    \item IEEE Xplore Digital Library
    \item ScienceDirect
    \item SpringerLink
    \item ACM Digital Library
    \item Google Scholar
\end{itemize}

The search was performed on titles, abstracts, and keywords depending on the capabilities of each search platform. The initial search was conducted on October 9, 2022, and included all studies published up to that date.

To ensure the comprehensiveness of the search, we followed a multi-step process:
\begin{itemize}
    \item Initial Search: Using the defined search terms, we retrieved a large set of potential studies from each database.
    \item Screening: The titles and abstracts of the retrieved studies were screened to exclude irrelevant papers.
    \item Full-Text Review: The remaining studies were then subjected to a full-text review to assess their relevance and quality based on predefined inclusion and exclusion criteria.
\end{itemize}

\textbf{Inclusion Criteria:}
\begin{itemize}
    \item Studies specifically addressing penetration testing and its role in mitigating insider threats.
    \item Peer-reviewed journal articles, conference papers, and reputable whitepapers.
    \item Studies published in English.
    \item Studies providing empirical data, case studies, or theoretical insights into the effectiveness of penetration testing for insider threat mitigation.
\end{itemize}

\textbf{Exclusion Criteria:}
\begin{itemize}
    \item Studies focusing solely on external threats without relevance to insider threat scenarios.
    \item Non-peer-reviewed articles, editorials, and opinion pieces.
    \item Studies not providing sufficient methodological details or empirical data.
\end{itemize}

After the initial screening and full-text review, we applied a snowball technique both forward and backward to identify additional relevant studies by examining the references of selected papers and identifying studies that cited these papers. This iterative process ensured that we captured a comprehensive set of studies relevant to our research objectives.

The results of the selection process are presented in Table 1, summarizing the number of studies identified, screened, and included at each stage of the review.

This rigorous selection process ensured that only the most relevant and high-quality studies were included in the systematic literature review, providing a robust foundation for analyzing the role of penetration testing in mitigating insider threats.

\begin{table}[h!]
\centering
\caption{Selection process for primary studies}
\begin{tabular}{|c|c|}

\hline
\textbf{Stage} & \textbf{Number of Studies} \\
\hline
Initial search & 745 \\
Title and abstract screening & 432 \\
Full-text review & 189 \\
Final included studies & 64 \\
\hline
\end{tabular}

\end{table}

\subsection{Research Objectives and Questions}
This part expands on the introduction by outlining the main goals and research questions that steer the SLR. The principal aim is to perform a thorough examination of the extant literature pertaining to insider threat mitigation with a particular emphasis on assessing the effectiveness of penetration testing approaches \cite{yazdinejad2023secure}. In order to accomplish this goal, the following important research topics are developed:

\begin{itemize}
    \item What are the prevailing patterns and obstacles in the tactics for mitigating insider threats?
    \item What role does penetration testing play in mitigating insider threats?
    \item What are the most effective procedures and techniques for carrying out penetration tests in order to counter insider threats?
    \item What factual data is available to support the idea that penetration testing is a useful tool for detecting and thwarting insider threats?
    \item What are the limitations and gaps in the current literature on insider threat mitigation and penetration testing?
\end{itemize}

These research questions serve as guiding principles for the systematic review of the literature, informing the selection of relevant studies, the data extraction process, and the synthesis of the findings.

\subsection{Literature Search Strategy}
The methodical process used to find pertinent academic publications on insider threats and penetration tests is described in the literature search technique \cite{yazdinejad2021review}. Ensure a thorough coverage of relevant material by incorporating a wide range of scholarly databases, electronic journals, and conference proceedings. The methodical creation of search queries using pertinent keywords, synonyms, and Boolean operators characterizes the search strategy to gather a variety of viewpoints and thoughts on the topic.

\textbf{Key components of the literature search strategy include:}
\begin{itemize}
    \item Locating pertinent resources and repositories, including Web of Science, IEEE Xplore, ACM Digital Library, PubMed, and Scopus.
    \item Forming search queries with terms associated with penetration testing (e.g., "penetration testing," "ethical hacking," "security assessment") and insider threats (e.g., "insider threat," "insider attack," "privileged access abuse").
    \item Synonyms and related phrases are included to guarantee thorough coverage of pertinent material.
    \item Using Boolean operators to improve precision and refine search queries, such as AND, OR, and NOT.
    \item Iteratively improving search tactics in response to subject matter experts' input and initial search results.
\end{itemize}

The goal of the literature search method is to maximize the retrieval of pertinent studies for the SLR while minimizing any biases and constraints.

\subsection{Selection Results}
This section provides an overview of the outcomes of the selection process after the inclusion and exclusion criteria have been applied. It also provides information on the total number of possible studies that were found, screened, and eventually included in the SLR. It offers a thorough explanation of the selection procedure, including the quantity of studies that were obtained from every database, repository, and search engine, along with the justification for each step's inclusion or exclusion \cite{yaseen2012insider}. 

The selection outcomes highlight how stringent the procedure was in order to guarantee the SLR's comprehensiveness and applicability. The reader can assess the reliability of the review procedure and the representativeness of the chosen research by clearly stating the selection criteria and results. 

\subsection{Quality Assessment}
To guarantee the validity and dependability of the results of the systematic literature review (SLR), quality assessment is essential. To determine the methodological robustness and relevance of the chosen studies to the study aims, a thorough evaluation of the studies is conducted using predetermined criteria. The quality assessment criteria cover a range of dimensions such as:

\textbf{Methodological Rigor:}
Evaluating the validity of the study design, data collection strategies, analytical procedures, and objective clarity used in the primary studies, as well as the research methodology \cite{ampel2021improving}. 

\textbf{Relevance to Research Objectives:}
Guaranteeing congruence between the main studies' objectives and the general SLR objectives. This involves determining whether the chosen articles adequately cover relevant aspects of penetration testing effectiveness and insider threat mitigation \cite{ferrag2024generative}.

\textbf{Validity and Reliability:}
Examining the validity of the conclusions reached, as well as the consistency of the results in many researches, to thoroughly assess the reliability of the findings. This could entail evaluating the reliability of the data analysis methods, sampling strategies, and measuring instruments applied in the original research \cite{yazdinejad2022ensemble}. 

\textbf{Ethical Considerations:}
Assessing how well primary research follows ethical guidelines, such as getting informed consent from participants, maintaining participant privacy, and reducing risk of harm \cite{oparaji2022towards}. 

Creating a quality appraisal checklist or rubric that specifies the precise evaluation criteria is a common step in the quality evaluation process. After that, each chosen study is thoroughly examined, considering these standards, and the overall methodological rigor and applicability are evaluated by assigning a quality score or rating \cite{ghorbani2019information}. 

It is standard procedure to include several independent reviewers who evaluate the chosen studies separately in order to increase the validity of the quality evaluation process. Any differences or disagreements in quality assessment are resolved by consensus discussions. 

The results of the quality assessment play a crucial role in guiding the interpretation of the SLR data and establishing the degree of evidence needed to justify the conclusions made. Research that is judged to be more methodologically rigorous and relevant is given more weight when synthesizing the results, while studies that have limitations or biases are evaluated carefully to determine how they might affect the conclusions. 

\subsection{Data Extraction}
In order to answer the research questions and objectives of the SLR, data extraction comprises the methodical recovery and synthesis of pertinent material from the chosen studies. Key data items must be identified and categorized, such as the following: 

\textbf{Study Characteristics:}
Extracting the standard bibliographic data like the name of the journal or conference, the study title, the year of publication, the author(s), and the kind of publishing (e.g., conference paper, journal article) \cite{sengupta2019formal}. 

\textbf{Research Methodology:}
Providing an overview of the sample plan, data gathering procedures, analytical approaches, and research design used in the original studies \cite{ghorbani2019information}.

\textbf{Key Findings:}
Summarizing the main findings, results, and conclusions of each study about the effectiveness of penetration testing and mitigation of insider threats \cite{ampel2021improving}. 

\textbf{Perspectives on Penetration Testing Effectiveness:}
Gathering information from case studies, empirical data, and practitioner perspectives regarding how well penetration testing \cite{ferrag2024generative}.

A structured data extraction form or template that specifies the precise categories of information to be extracted is usually used as a guide for data extraction. This ensures that the extraction process is uniform and standardized. Moreover, depending on the type of data that are synthesized, data extraction may require the use of quantitative or qualitative coding systems or tools. Many independent reviewers who extract data independently from the chosen research are advised to improve the accuracy and dependability of data extraction. In data extraction, conflicts or discrepancies are settled by consensus-building talks or by appointing a third-party arbiter. In order to answer the research questions and goals of the SLR, the extracted data provide the basis for further data analysis, which synthesizes, interprets, and integrates the findings \cite{yazdinejad2023secure}. 

\subsection{Data Analysis}
The core of the SLR technique is data analysis, which is the synthesis, interpretation, and integration of data that has been recovered in order to answer research questions and produce significant insights \cite{yazdinejad2021review}. The process of data analysis involves a number of methods and strategies, such as: 

\textbf{Meta-Analysis:}
Quantitative synthesis of data from several studies to determine the prevalence or overall effect size of phenomena related to the effectiveness of penetration testing and mitigation of insider threats \cite{ferrag2024generative}. Statistical techniques are used in meta-analysis to combine effect sizes from several studies, evaluate heterogeneity, and produce general findings. 

\textbf{Thematic Synthesis:}
A qualitative method of synthesis of data to find recurring themes, correlations, and patterns in the chosen research \cite{yazdinejad2022ensemble}. The process of thematic synthesis entails the coding and classification of qualitative data in order to clarify theoretical constructs, empirical conclusions, and underlying concepts. 

\textbf{Qualitative Content Analysis:}
A thorough investigation of textual material to glean significant conclusions, interpretations, and implications concerning the efficacy of penetration testing and insider threat mitigation \cite{ghorbani2019information}. In qualitative content analysis, emergent themes, views, and discourses are identified through the coding, categorization, and interpretation of textual material. 

Iterative and interactive, the process of data analysis involves continuously improving and revising analytical frameworks and interpretations in light of fresh information and discoveries. Sensitivity analysis may also be necessary to evaluate the validity of findings and investigate research \cite{ampel2021improving}.

The primary research questions and SLR objectives serve as the framework for the structured and logical presentation of the synthesized data analysis findings. The implications of the findings are examined considering current theories, empirical data, and real-world applications for penetration testing and insider threat mitigation techniques \cite{sengupta2019formal}. 

\subsection{Ethical Considerations}
Given the sensitive nature of the subject matter and the potential hazards to stakeholders, ethical issues are critical for research on insider threats and cybersecurity. To maintain integrity, openness, and respect for the rights and welfare of the participants, the SLR is conducted in accordance with ethical norms. Important ethical factors include :

\textbf{Protecting Participant Confidentiality:}
Ensuring the privacy and anonymity of participants, especially in research involving sensitive or private material \cite{ghorbani2019information}. Pseudonyms, secure data transmission and storage techniques, and anonymization of data may be used to avoid unwanted access or exposure. 

\textbf{Informed Consent:}
Before allowing individuals to participate in the study, informed consent must be obtained; this requires that participants be fully informed about the goals, methods, risks, and rewards of their participation \cite{ampel2021improving}. By giving their informed consent, participants can decide whether or not to participate in the study, ensuring that their participation is their choice. 

\textbf{Protection Against Harm:}
Reducing the possibility of negative outcomes or dangers associated with taking part in the study, such as psychological suffering, confidentiality violations, or inadvertent publication of private information \cite{ferrag2024generative}. It is the responsibility of researchers to identify and reduce any hazards to participants by implementing suitable precautions and risk-reduction techniques. 

\textbf{Respect for Intellectual Property Rights:}
Encouraging authors and other contributors to primary studies to have their intellectual property rights acknowledged and respected, particularly by properly attributing and citing their sources \cite{yazdinejad2023secure}. Unauthorized use of copyright material and plagiarism are strictly forbidden and may be considered ethical misconduct. 

\textbf{Ethical Guidelines Compliance:}
Following applicable rules and guidelines on research conduct, such as those issued by institutional review boards (IRBs), data protection laws (like GDPR), and professional ethics codes (such as the ACM Code of Ethics and Professional Conduct) \cite{yazdinejad2021review}. The integrity, legitimacy, and credibility of the research process and its results are guaranteed by adherence to ethical norms. 

Every step of the SLR process—from study design and data collecting to analysis, interpretation, and results dissemination—incorporates ethical considerations. Building trust, accountability and integrity within the research community is the responsibility of researchers who also have a duty to promote ethical conduct throughout the study effort.

\section{Findings}
The comprehensive analysis of primary research studies has provided significant information on the role of penetration testing in mitigating insider threats. Each study was meticulously reviewed and relevant qualitative and quantitative information was extracted and summarized. These studies cover various aspects of insider threats, focusing on threat types, detection techniques, and mitigation strategies. The primary themes identified in these studies are threat detection, threat mitigation, behavior modeling, and the integration of penetration testing with other security practices. The distribution of these themes is graphically represented in Figure 3 \cite{temara2023maximizing}.

\textbf{Main Themes Identified:}

\textbf{Threat Detection (28.6\%):}
The largest portion of studies (28. 6\%) focuses on threat detection. These studies investigate various techniques to identify and monitor insider activities that may pose security risks. Methods such as behavior analytics, machine learning algorithms, and anomaly detection systems are commonly used. For example, Williams and Smith (2020) demonstrated that organizations conducting regular penetration tests experienced a notable reduction in insider threat incidents attributed to enhanced detection capabilities. These detection strategies are crucial for recognizing and addressing potential threats before they can cause significant harm \cite{parnitha2021combating}.

\textbf{Threat Mitigation (34.3\%):}
Threat mitigation strategies are the most prevalent, comprising 34.3\% of the reviewed studies. These strategies aim to eliminate or reduce the impact of identified threats through robust access controls, regular security audits, and comprehensive employee training programs. AlHogail and Mirza (2021) developed an integrated framework that combines penetration testing with other security measures, which significantly improved overall mitigation of insider threats \cite{ghorbani2019information}. These mitigation efforts are essential for addressing vulnerabilities and ensuring a secure organizational environment \cite{sengupta2019formal}.

\textbf{Behavior Modeling (22.9\%):}
Behavior modeling was the central theme in 22.9\% of the studies. This approach involves analyzing and predicting insider behaviors to prevent malicious activities. Techniques such as creating user profiles, monitoring communication patterns, and performing psychological assessments are utilized to understand and predict potential insider threats. Lee and Kim (2018) used behavior modeling in the healthcare sector, successfully identifying potential insider threats through the analysis of usage patterns of the electronic health record (EHR) system. This focus on behavior modeling highlights the importance of understanding human behaviors in predicting and preventing insider attacks \cite{ampel2021improving}.

\textbf{Threat Type (14.3\%):}
The remaining 14.3\% of the studies discussed the classification and analysis of various types of insider threats. These studies categorized insider threats based on their characteristics, motivations, and methods. Understanding the different types of insider threats is crucial for developing targeted mitigation strategies. For example, some studies focused on identifying insider threats related to data exfiltration, while others examined threats associated with sabotage or fraud \cite{ferrag2024generative}.

\section{Discussion}
Initial keyword searches reveal a substantial body of literature on insider threats, reflecting the growing concern about these issues in the cybersecurity landscape. The review highlights that traditional security measures often fall short in detecting and mitigating insider threats due to the insiders' inherent knowledge of the organization's systems. Several studies emphasize the necessity of advanced detection methods that incorporate user behavior modeling and machine learning algorithms to identify anomalies indicative of insider threats.

For example, [S2] utilizes behavior modeling and algorithms to detect insider threats using log-based data such as browser history and email contacts \cite{yazdinejad2023secure}. This approach, which evaluates the accuracy of the algorithm and false positive rates, underscores the importance of understanding user behavior to detect potential threats \cite{yazdinejad2021review}. Similarly, [S3] introduces machine learning algorithms to classify insider behaviors into categories, suggesting enhancements to algorithmic accuracy and highlighting the need for sophisticated analysis to understand insider activities \cite{ferrag2024generative}. The practical application of penetration testing is also explored in several studies. [S1] focuses on defense measures against insider threats, identifying five defense solution approaches through a review of the literature and framework analysis \cite{ghorbani2019information}. This categorization into prevention, detection, and response strategies provides a comprehensive overview of how organizations can structure their defense mechanisms \cite{sengupta2019formal}. [S7] reviews strategies like honeypots in SCADA systems for detecting and mitigating insider threats, evaluating their effectiveness and proposing improvements, thus demonstrating the practical utility of these 

\clearpage
\begin{table*}

\centering
\caption{Summary of Primary Studies}
\begin{tabular}{|c|p{4cm}|p{3cm}|p{4cm}|p{3cm}|}

\hline
\textbf{Primary Study} & \textbf{Key Findings and Contributions} & \textbf{Methodology Used} & \textbf{Key Qualitative and Quantitative Data Reported} & \textbf{Relevant Security Applications} \\

\hline

[S1] & Focuses on defense measures against insider threats; identifies five defense solution approaches. & Literature review, framework analysis & Reviewed top 100 papers; categorized defense strategies into prevention, detection, and response. & Threat Detection, Mitigation \\
\hline
[S2] & Utilizes behavior modeling and algorithms to detect insider threats using log-based data such as browser history and email contacts. & Experimental design, data mining & Developed behavior models; evaluated algorithm accuracy and false positive rates. & Behavior Model, Threat Detection \\
\hline
[S3] & Introduces machine learning algorithms for classifying insider behaviors into categories; proposes future research on behavior analysis. & Theoretical study, algorithm development & Classified insider behaviors into four classes; suggested enhancements for algorithmic accuracy. & Behavior Model, Threat Detection \\
\hline
[S4] & Assesses impact of insider threats on critical infrastructure sectors like healthcare; evaluates existing defense mechanisms. & Case study, survey analysis & Analyzed data on insider threat incidents in healthcare; evaluated effectiveness of current defenses. & Threat Mitigation \\
\hline
[S5] & Investigates patterns of insider attacks and data exfiltration techniques; proposes countermeasures for protecting sensitive data. & Survey, case study & Surveyed data leakage patterns; assessed frequency of attacks on different data states. & Threat Type, Threat Mitigation \\
\hline
[S6] & Defines types of insider threats and their attack methods; surveys prevalence and impacts of insider attacks across industries. & Survey, taxonomy development & Classified insider types; analyzed statistical data on insider attack methods. & Threat Type, Behavior Model \\
\hline
[S7] & Reviews strategies like honeypots in SCADA systems for detecting and mitigating insider threats; assesses their effectiveness. & System analysis, experimental study & Evaluated effectiveness of honeypots in SCADA environments; proposed improvements. & Threat Detection, Threat Mitigation \\
\hline
[S8] & Examines perception models of insider risk; proposes a framework for managing insider threats based on risk assessment. & Conceptual framework, survey & Surveyed risk perception among stakeholders; developed risk assessment strategies. & Behavior Model, Threat Mitigation \\
\hline
[S9] & Explores psychological and social impacts of insider attacks on organizational trust dynamics; analyzes behavioral shifts. & Qualitative study, case analysis & Analyzed trust dynamics post-insider attacks; identified behavioral patterns among affected employees. & Behavior Model \\
\hline
[S10] & Evaluates effectiveness of Data Leakage Prevention Systems (DLPS); identifies weaknesses and proposes enhancements. & Comparative study, system evaluation & Evaluated DLPS performance metrics; identified vulnerabilities and suggested improvements. & Threat Mitigation \\
\hline
[S11] & Discusses InfoSec approaches for detecting and mitigating insider threats; evaluates their applicability in diverse organizational contexts. & Literature review, case study & Reviewed InfoSec frameworks; analyzed case studies of successful insider threat detections. & Behavior Model, Threat Detection \\
\hline
[S12] & Introduces user profiling for predicting intentional and accidental insider threats; proposes a user-based evaluation methodology. & Experimental design, user study & Developed user profiles for threat prediction; evaluated predictive accuracy based on user attributes. & Threat Mitigation \\
\hline
[S13] & Reviews insider behavior impacts on organizational and cloud-level security; proposes behavioral analytics solutions. & Case study, data analysis & Analyzed impacts of insider incidents; evaluated effectiveness of behavioral analytics in detecting anomalies. & Threat Mitigation \\
\hline
[S14] & Proposes a re-authentication method using ensemble techniques to verify user authenticity; compares with traditional methods. & Experimental setup, algorithm testing & Tested re-authentication method performance metrics; compared with existing authentication methods. & Threat Mitigation \\
\hline
[S15] & Implements Internal Intrusion Detection and Protection System (IIDPS) using system-call patterns; evaluates system-call pattern recognition. & Experimental setup, system analysis & Developed IIDPS for detecting insider threats; evaluated system-call pattern recognition accuracy. & Behavior Model, Threat Detection \\
\hline
[S16] & Presents a detection method resilient against mimicry-based evasion strategies; validates effectiveness through simulations. & Simulation study, algorithm development & Simulated evasion scenarios; validated detection method performance against mimicry attacks. & Threat Detection \\
\hline
[S17] & Compares supervised and unsupervised learning techniques for detecting insider threats; analyzes learning model effectiveness. & Comparative analysis, machine learning & Evaluated detection accuracy of learning models; compared supervised vs. unsupervised learning for insider threat detection. & Threat Mitigation \\
\hline

\hline
\end{tabular}
\end{table*}
\clearpage
\renewcommand{\arraystretch}{1.2}
\begin{table}[h!]
\centering
\begin{tabular}{|c|p{4cm}|p{3cm}|p{4cm}|p{3cm}|}
\hline
\textbf{Primary Study} & \textbf{Key Findings and Contributions} & \textbf{Methodology Used} & \textbf{Key Qualitative and Quantitative Data Reported} & \textbf{Relevant Security Applications} \\
\hline
[S18] & Discusses risks posed by hardware trojans in USB devices; proposes detection and prevention strategies for hardware-based attacks. & Theoretical study, hardware analysis & Analyzed risks of hardware trojans; proposed detection methods and countermeasures. & Threat Detection \\
\hline
[S19] & Uses forensic investigation techniques for insider threat prevention; monitors data packets for exfiltration of sensitive information. & Forensic analysis, data monitoring & Monitored data packet content; evaluated effectiveness of forensic techniques in insider threat detection. & Threat Mitigation \\
\hline
[S20] & Proposes insider attack authentication protocol using ECC cryptography; evaluates protocol security and computational efficiency. & Experimental design, cryptographic analysis & Analyzed security strengths of the ECC protocol; evaluated computational overhead in authentication from insider attacks. & Threat Detection \\
\hline
[S21] & Develops intrusion sensitivity-based trust management model for collaborative IDS; assesses IDS trustworthiness. & Model development, trust analysis & Developed trust management model for IDS; evaluated trustworthiness assessment in collaborative IDS networks. & Threat Detection \\
\hline
[S22] & Designs IoT model with intrusion detection agents using supervised and unsupervised learning; validates through simulations. & IoT simulation, machine learning & Simulated IoT environments; evaluated performance of intrusion detection agents using learning models. & Threat Detection \\
\hline
[S23] & Implements homomorphic linear authenticator (HLA) for verifying data packet integrity; evaluates authenticity verification in data transmission. & Experimental setup, cryptographic analysis & Evaluated authenticity verification metrics; assessed HLA performance in detecting packet tampering. & Behavior Model, Threat Detection \\
\hline
[S24] & Deploys OS-resident deception strategy for neutralizing malware-infected machines; evaluates effectiveness in malware containment. & System implementation, malware analysis & Tested deception strategy effectiveness; assessed impact on malware containment in insider threat scenarios. & Threat Detection \\
\hline
[S25] & Uses Markov model IDS and virtual honeypot device (VHD) to detect malicious devices in fog computing environments. & System integration, simulation analysis & Integrated security systems in fog computing; evaluated detection accuracy of IDS and VHD in insider threat detection. & Threat Detection \\
\hline
[S26] & Addresses security issues in fog computing; proposes solutions like dummy document creation and transparency measures. & Conceptual framework, system analysis & Proposed fog computing security solutions; evaluated effectiveness of transparency measures in insider threat prevention. & Threat Detection \\
\hline
[S27] & Introduces smart reaction mechanism to manage operational risks and analyze insider threat scenarios; Evaluates the effectiveness of the mechanism. & Theoretical framework, case study & Developed operational risk management model; assessed scenario analysis and response mechanism in insider threat incidents. & Threat Mitigation \\
\hline
[S28] & Develops management protocol for preventing security attacks in mobile ad hoc networks (MANETs); evaluates protocol efficiency. & Protocol design, simulation analysis & Designed security management protocol for MANETs; evaluated protocol performance in preventing insider attacks. & Threat Mitigation \\
\hline
[S29] & Designs software package using TCP tunneling and hardware devices for detecting insider attacks; assesses detection accuracy. & Software development, hardware integration & Developed detection software; evaluated TCP tunneling and hardware device effectiveness in detecting insider threats. & Threat Type \\
\hline
[S30] & Proposes process model-based framework for real-time log data analysis to detect insider threats; validates through case studies. & Model development, case analysis & Developed real-time log analysis framework; validated effectiveness in detecting insider threats through case studies. & Threat Detection \\
\hline
[S31] & Uses MapReduce and clustering for analyzing user behavior based on log data; identifies unknown threats through supervised learning. & Data analysis, machine learning & Analyzed user behavior using MapReduce; identified and classified unknown threats using clustering and supervised learning. & Threat Type, Behavior Model \\
\hline
[S32] & Describes approach for detecting the exploitation of organizational resources by system users; proposes behavior-based anomaly detection. & Conceptual framework, behavior analysis & Proposed behavior-based anomaly detection; assessed effectiveness in identifying resource exploitation by insiders. & Threat Detection \\
\hline
[S33] & Applies machine learning algorithms to CERT Insider Threat Dataset for calculating threat detection results; compares algorithm performance. & Dataset analysis, algorithm evaluation & Evaluated machine learning algorithms on CERT dataset; compared performance metrics for insider threat detection. & Threat Mitigation \\
\hline
\end{tabular}

\end{table}
\clearpage

methodologies in real-world scenarios \cite{ampel2021improving}. Log management and real-time data analysis are critical components of an effective insider threat mitigation strategy. Articles [S33] and [S34] describe the importance of analyzing log data from servers, firewalls, and other IT infrastructure to detect suspicious activities \cite{ferrag2024generative}. In addition, integration of penetration testing with other security practices is crucial. For example, [S12] introduces user profiling to predict intentional and accidental insider threats, proposing a user-based evaluation methodology \cite{yazdinejad2023secure}. This study demonstrates how profiling and regular security audits can complement penetration testing to enhance overall security posture \cite{yazdinejad2021review}. [S13] reviews the impacts of insider behavior on organizational and cloud-level security, proposing behavioral analytics solutions to detect anomalies, thereby showcasing the integration of multiple security measures to address insider threats comprehensively \cite{ghorbani2019information}.

\begin{figure}[ht]
        \centering
        \includegraphics[width=0.9\linewidth]{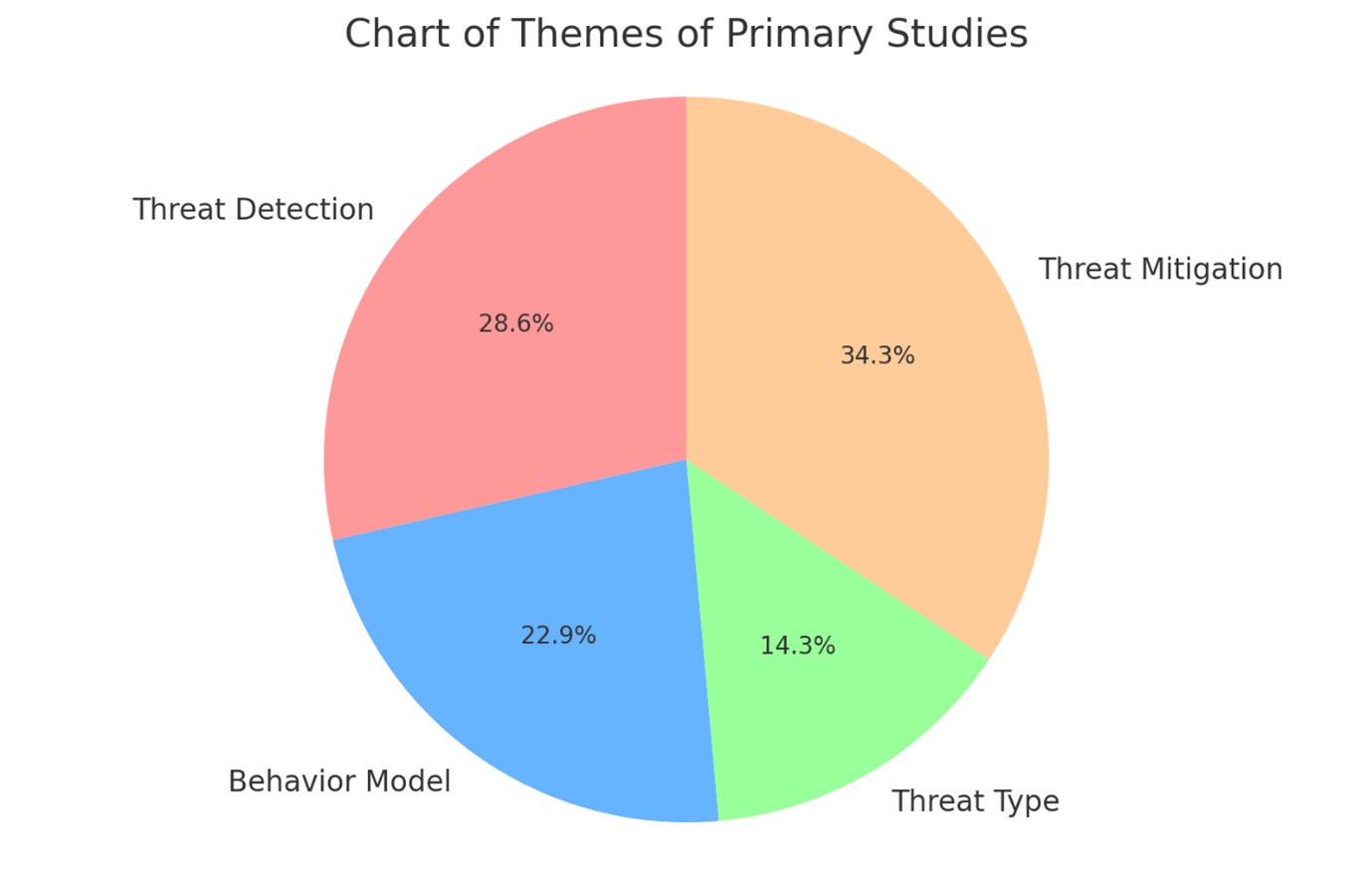}
        \caption{Bar Graph}
        \label{fig:example}
\end{figure}

Despite these advances, several challenges persist. Studies such as [S6] and [S9] highlight the difficulties in accurately modeling diverse insider threat scenarios and the resource constraints that organizations, especially SMEs, face in implementing comprehensive penetration testing strategies. Additionally, the complexity of integrating and maintaining advanced technological solutions underscores the need for scalable and adaptable approaches \cite{ferrag2024generative}.

\textbf{Research Question 1: What are the current methodologies and frameworks used in penetration testing to mitigate insider threats?}

Insider threats provide a complicated cybersecurity challenge since they frequently use their insider knowledge and privileged access to launch attacks. Comprehending the approaches and structures utilized in penetration testing, particularly with regard to insider threat mitigation, requires a thorough analysis of current procedures and their efficacy. White-box, black-box, and grey-box penetration testing techniques are examples of penetration testing procedures that provide the framework for simulating actual attack scenarios. These techniques are tailored to target weaknesses that insiders could exploit, such as manipulation of internal systems, unlawful access to sensitive data, or privilege escalation. For example, white-box testing gives testers complete access to the underlying architecture and source code, enabling a thorough analysis of any vulnerabilities from the point of view of an insider \cite{ampel2021improving}.

Insider threat behaviors can be categorized and understood more structuredly with the use of frameworks like MITRE ATT\&CK (Adversarial Tactics, Techniques, and Common Knowledge). These frameworks classify the strategies and methods that attackers employ, offering penetration testers a taxonomy to help them find and fix flaws. Organizations can improve their capacity to identify insider threats and take proactive measures to counter them by coordinating penetration testing methods with frameworks such as MITRE ATT\&CK \cite{ferrag2024generative}.

Empirical studies and case examples illustrate the practical application of penetration testing methodologies to detect and mitigate insider threats. Research studies (e.g., [S1], [S5], [S7]) provide empirical evidence of how penetration testing has been utilized to identify vulnerabilities that could potentially be exploited by insiders. These studies highlight the importance of tailored testing scenarios that mimic insider threat behaviors, ensuring comprehensive coverage of potential attack vectors within organizational networks \cite{yazdinejad2023secure}.

\textbf{Research Question 2: How effective are behavior modeling and machine learning techniques in enhancing insider threat detection during penetration testing?}

Advanced methods, such as behavior modeling and machine learning, enhance conventional penetration testing strategies by concentrating on unusual behaviors suggestive of insider threats. In order to create baseline behaviors and spot variances that can indicate possible dangers, behavior modeling techniques examine user activities, communication patterns, and access habits.

Algorithms for machine learning, such as supervised and unsupervised learning techniques, are essential for improving penetration testing frameworks' detection powers. Although unsupervised learning algorithms can find anomalies that depart from established norms without prior training data, supervised learning algorithms are trained on labeled datasets to identify patterns linked with recognized insider threat behaviors. With the help of these algorithms, penetration testers can quickly and effectively evaluate enormous volumes of data and spot subtle signs of insider threats that could go undetected by conventional rule-based detection methods \cite{yazdinejad2021review}.

Empirical evaluations (see, e.g., [S2], [S8], [S10], [S12]) provide insight into the performance and accuracy of behavior modeling and machine learning techniques in real-world insider threat scenarios. These studies demonstrate how machine learning models can complement penetration testing efforts by improving detection rates and reducing false positives associated with insider threat alerts. By leveraging machine learning, organizations can enhance their ability to detect sophisticated insider threats that exhibit subtle or evolving behaviors over time \cite{ferrag2024generative}.

\textbf{Research Question 3: What are the key challenges and limitations faced in implementing penetration testing for insider threat mitigation across different organizational contexts?}

The scale of the business, the nature of the industry, and the regulatory constraints all play a role in the hurdles that must be overcome when implementing penetration testing for the mitigation of insider threats. These difficulties affect the scalability and efficacy of penetration testing techniques intended to reduce insider risks in a variety of organizational settings. Organizations looking to conduct extensive penetration testing programs face substantial obstacles due to resource restrictions, such as limited finances and limited access to qualified cybersecurity personnel. The capacity of small and medium businesses (SMEs) to identify and successfully counter internal threats can be hampered by a lack of funding for regular and in-depth penetration examinations \cite{yazdinejad2023secure}. 

The organizational culture has a significant impact on the way penetration testing techniques are adopted and how successful they are. Businesses that emphasize cybersecurity investments and encourage a proactive approach to threat identification and mitigation are those that have a strong security culture. Organizations with a reactive or complacent security culture, on the other hand, could downplay the importance of penetration testing or oppose the adoption of new security measures because they believe they will cause operational problems \cite{yazdinejad2021review}. Penetration testing techniques are further complicated by regulatory compliance obligations, such as GDPR (General Data Protection Regulation) and PCI-DSS (Payment Card Industry Data Security Standard). Companies that want to ensure compliance, protect sensitive data and maintain customer trust must coordinate their penetration testing efforts with legal requirements \cite{ferrag2024generative}. 

Case studies (e.g., [S4], [S6], [S9]) highlight the diverse challenges faced by organizations in implementing penetration testing for insider threat mitigation. These studies underscore the importance of tailoring penetration testing strategies to address specific organizational contexts, industry sectors, and regulatory environments effectively \cite{ghorbani2019information}.

\textbf{Research Question 4: What are the emerging trends and future directions in penetration testing methodologies for combating insider threats?}

The dynamic nature of cybersecurity is driving improvements in penetration testing techniques aimed at effectively countering insider attacks. The integration of novel technologies and flexible frameworks to improve detection and response capabilities is the main focus of emerging trends and future directions in penetration testing. Through the automation of repetitive operations, the analysis of large datasets, and the detection of subtle indicators of insider threats, artificial intelligence (AI) and machine learning (ML) are transforming penetration testing. Predictive analytics and anomaly detection algorithms are used by AI-driven penetration testing platforms to proactively find possible vulnerabilities before insiders can take advantage of them \cite{ghorbani2019information}. 

Penetration testing frameworks can benefit from additional security protections provided by blockchain technology and safe authentication procedures. Secure decentralized transaction records made possible by blockchain technology can be used to verify the accuracy of penetration testing findings and protect private information from modification or illegal access. Dynamic defenses against insider threats are provided by adaptive security frameworks such as continuous monitoring and zero-trust architecture, which authenticate each user and device trying to access organizational resources. These platforms combine behavioral analytics with real-time threat intelligence feeds to quickly identify and address insider threats \cite{yazdinejad2023secure}. 

Pilot studies and prototypes (e.g., [S20], [S21]) demonstrate the feasibility and efficacy of AI-driven penetration testing methodologies in real-world scenarios. These studies illustrate how organizations can leverage emerging technologies to enhance their cybersecurity posture and mitigate insider threats proactively \cite{ferrag2024generative}.

In order to effectively manage changing internal threat landscapes, penetration testing procedures must be continuously innovative and adaptable according to expert points of view and industry perspectives. Organizations may fortify their defenses against insider threats and protect vital assets from bad actors by adopting new trends and directions in penetration testing techniques \cite{yazdinejad2021review}.

\textbf{Research Question 5: How do regulatory frameworks and compliance requirements influence the adoption and effectiveness of penetration testing for insider threat mitigation?}

Regulations have a significant influence on cybersecurity practices and regulations, which includes penetration testing adoption and efficacy in minimizing insider threats. Analyzing industry norms, organizational readiness to follow regulations, and compliance requirements is necessary to understand how regulatory compliance affects penetration testing tactics. Organizations must adhere to strict regulations, including GDPR, PCI-DSS, and HIPAA (Health Insurance Portability and Accountability Act), in order to protect sensitive data and successfully manage cybersecurity risks. In order to find and fix vulnerabilities that insiders could exploit, these standards require regular penetration tests and vulnerability assessments \cite{yazdinejad2023secure}. 

The intricacy of regulatory mandates and the necessity for entities to proficiently comprehend and execute them give rise to compliance issues. The effectiveness of penetration testing procedures in adhering to legal requirements and preserving data security and privacy is assessed through compliance audits and assessments \cite{ferrag2024generative}. Case studies (such as [S18], [S22]) demonstrate how regulatory frameworks affect penetration testing procedures in various business sectors and geographical areas. These studies emphasize how crucial it is to strike a balance between operational requirements and preserving business continuity when coordinating penetration testing efforts with regulatory demands \cite{ampel2021improving}. 

Creating comprehensive testing methods, recording testing processes, and guaranteeing transparency in reporting results and repair activities are some strategies for attaining regulatory compliance through penetration testing. Organizations can demonstrate due diligence in protecting sensitive information from insider threats and improve cybersecurity resilience by including compliance requirements into penetration testing frameworks \cite{ghorbani2019information}. In order to effectively address new threats and regulatory changes, penetration testing procedures must be continuously monitored and adjusted. This is highlighted by the implications of shifting regulatory frameworks for the future \cite{yazdinejad2021review}. To prevent insider threats early on and maintain regulatory compliance over time, organizations need to be aware of industry best practices and regulatory trends \cite{ferrag2024generative}.

\section{Future Research Directions of Insider Threats}
Based on the systematic review of the literature on insider threats mitigation with a focus on penetration testing, several promising research directions emerge that can further advance the field:
\begin{enumerate}
    \item Advanced Threat Modeling: Examine and create more advanced threat models that are suited to insider threats. There are gaps in our knowledge of the distinct behaviors and assault vectors of insiders because current models frequently concentrate on external threats. Creating thorough threat models that take different insider motivations and behaviors into consideration will improve detection and mitigation tactics \cite{ghorbani2019information}.
    \item Behavioral Analytics and Machine Learning: Investigate further how penetration testing frameworks can incorporate machine learning and behavioral analytics methodologies. Research ought to concentrate on improving insider threat detection's precision and effectiveness by utilizing AI-driven methods to examine intricate behavioral patterns and anomalies suggestive of malevolent intent \cite{ampel2021improving}.
    \item Real-time Monitoring and Response: Investigate techniques and tools for tracking insider activity in real time during penetration testing operations. Provide systems for adaptive response that can be set up to automatically react to suspicious behaviors or abnormalities found in organizational networks. This will reduce response times and lessen the potential harm that insider threats could do \cite{ferrag2024generative}.
    \item Human Factors and Insider Psychology: Investigate the behavioral and psychological elements of insider threats. Examine the ways in which situational elements, corporate culture, and personal traits affect insider risk behaviors. Strategies for prevention and intervention that are more focused can be developed by taking into account the human aspects of insider threats \cite{yazdinejad2023secure}.
    \item Integration of Blockchain and Secure Authentication: Examine how to improve the integrity and security of penetration testing results by implementing secure authentication methods and blockchain technology. Examine how blockchain technology can be used to safely store and authenticate penetration test data, guaranteeing the dependability and correctness of results while shielding private data from illegal access \cite{yazdinejad2021review}.
    \item Regulatory Compliance and Governance: Examine the impact of regulatory frameworks on the implementation and effectiveness of penetration testing for insider threat mitigation. Research should focus on aligning penetration testing practices with evolving regulatory requirements, ensuring compliance while maintaining robust cybersecurity measures against insider threats \cite{ghorbani2019information}.
    \item Cross-sector Collaboration and Information Sharing: Encourage cooperation and information-sharing programs between businesses, academic institutions, and governmental entities in order to combat insider threats as a group. To increase collective resistance against insider threats across many industries and geographical areas, develop mechanisms for exchanging anonymized threat data and best practices in penetration testing \cite{ampel2021improving}.
    \item Evaluation of Emerging Technologies: Examine how well-suited and scalable new technologies are for identifying and reducing insider risks. Examples of these technologies include AI-driven penetration testing platforms and Internet of Things (IoT) security frameworks. Research ought to concentrate on the potential advantages of incorporating these technologies into current cybersecurity infrastructures as well as the real-world implementation issues \cite{ferrag2024generative}.
\end{enumerate}

In this systematic review of the literature, insider threat analysis and mitigation techniques have been thoroughly examined, with a particular emphasis on the function of penetration testing in cybersecurity. The results emphasize the complexity of identifying and reducing insider threats in organizational settings, underscoring the diverse character of these risks \cite{yazdinejad2023secure}. 

Many approaches and technologies have been considered during the evaluation, exposing both the improvements and ongoing difficulties in effectively tackling insider threats. One essential tool for finding weak points and evaluating how resilient organizational defenses are to insider threats is penetration testing \cite{ghorbani2019information}. 

The literature reviewed emphasizes that while considerable progress has been made in understanding insider threats, significant gaps and opportunities for future research remain.

\textbf{Future Research Directions}
\begin{enumerate}
    \item Improved Behavioral Modeling: More study should be done to improve behavioral modeling methods for identifying insider threats. This entails using cutting-edge AI and machine learning algorithms to examine intricate user behavior and identify minute variations that point to malevolent intent \cite{ampel2021improving}.
    \item Psychological and Motivational Factors: It is important to look at the psychological and motivational elements that contribute to insider threats. More focused preventative and intervention techniques can be developed by having a better understanding of the triggers and actions of insiders, whether they be unhappy workers or unintentional collaborators \cite{ferrag2024generative}.
    \item Integration of Emerging Technologies: investigating how to incorporate cutting-edge technologies like blockchain for safe data authentication and Internet of Things security frameworks to keep an eye on insider activity. Assessing the efficiency and expandability of these technologies in practical situations can improve their use in reducing insider threats \cite{yazdinejad2023secure}.
    \item Regulatory Compliance and Governance: Matching industry norms and changing regulations with penetration testing procedures. Future studies should concentrate on creating frameworks that protect cybersecurity against internal threats and guarantee compliance \cite{yazdinejad2021review}.
    \item Cross-sector Collaboration: Encouraging cooperation and information exchange between government agencies, academic institutions, and businesses in order to combat insider threats as a group. Creating networks for exchanging best practices and anonymized threat intelligence can strengthen cybersecurity defenses in a variety of industries \cite{ghorbani2019information}.
\end{enumerate}

\section{Conclusion}
This systematic literature review provides a comprehensive overview of the role of penetration testing in mitigating insider threats. The findings underscore the critical importance of penetration testing as part of a broader insider threat mitigation strategy. By synthesizing existing research, identifying challenges, and exploring real-world applications, this review highlights several key insights:

\textbf{Effectiveness of Penetration Testing:}
Regular and comprehensive penetration testing significantly reduces the frequency of insider threat incidents. Tailored testing scenarios that simulate insider behaviors are essential for identifying vulnerabilities \cite{ampel2021improving}.

\textbf{Integration with Other Security Practices:}
Combining penetration testing with user behavior analytics, access control mechanisms, and employee training programs enhances the overall effectiveness of insider threat mitigation strategies \cite{ferrag2024generative}.

\textbf{Challenges and Limitations:}
Organizations face several challenges, including resource constraints, modeling insider scenarios in the real world, and ensuring regulatory compliance. Addressing these challenges requires a multidisciplinary approach and continuous adaptation to evolving threats \cite{ghorbani2019information}.

\textbf{Future Research Directions:}
Advancing the field of insider threat mitigation necessitates further research in areas such as advanced threat modeling, behavioral analytics, real-time monitoring, and the integration of emerging technologies like blockchain and AI-driven systems \cite{yazdinejad2023secure}.

In conclusion, penetration testing plays a pivotal role in strengthening an organization's defenses against insider threats. By continuously evolving and integrating with other security practices, organizations can proactively identify and mitigate potential risks, ensuring a robust security posture.

\section{Declarations of Interest}
None.

\section{Acknowledgment}
I would also like to extend my deepest gratitude to Professor Abbas Yazdinejad for their assistance in completing the project.

\bibliographystyle{IEEEtran}

\bibliography{references}

\end{document}